\documentclass[useAMS, usenatbib, apjl]{emulateapj}
\usepackage{afterpage}
\usepackage{tensor}
\usepackage{graphicx}
\usepackage{subfigure}
\usepackage{amsmath, amssymb, amsfonts}

\newcommand \divE {\nabla\cdot\vec{E}}

\newcommand \curlE {\nabla \times \vec{E}}
\newcommand \curlB {\nabla \times \vec{B}}
\newcommand \E {\vec{E}}
\newcommand \B {\vec{B}}
\newcommand \J {\vec{J}}
\newcommand \rlc {R_{\rm LC}}

\newcommand \dd {\partial}
\newcommand \beq {\begin{equation}}
\newcommand \eeq {\end{equation}}

\newcommand \lp {\left(}
\newcommand \rp {\right)}
\newcommand \lb {\left\{}
\newcommand \rb {\right\}}
\newcommand \ls {\left[}
\newcommand \rs {\right]}

\newcommand \rstar {r_{\star}}
\newcommand \thpc {\theta_{\rm pc}}
\newcommand \throt {\theta_{\rm rot}}
\newcommand \ftw {f_{\rm pc}}
\newcommand \frot {f_{\rm rot}}
\newcommand \tmax {t_{\max}}
\newcommand \Etw {E_{\rm tw}}

\let\oldhat\hat
\renewcommand{\vec}[1]{\boldsymbol{#1}}
\renewcommand{\hat}[1]{\oldhat{\mathbf{#1}}}

\def\bB{{\mathbf B}}
\def\RLC{R_{\rm LC}}

\newbox\grsign \setbox\grsign=\hbox{$>$} \newdimen\grdimen \grdimen=\ht\grsign
\newbox\simlessbox \newbox\simgreatbox \newbox\simpropbox
\setbox\simgreatbox=\hbox{\raise.5ex\hbox{$>$}\llap
     {\lower.5ex\hbox{$\sim$}}}\ht1=\grdimen\dp1=0pt
\setbox\simlessbox=\hbox{\raise.5ex\hbox{$<$}\llap
     {\lower.5ex\hbox{$\sim$}}}\ht2=\grdimen\dp2=0pt
\setbox\simpropbox=\hbox{\raise.5ex\hbox{$\propto$}\llap
     {\lower.5ex\hbox{$\sim$}}}\ht2=\grdimen\dp2=0pt
\def\simgt{\mathrel{\copy\simgreatbox}}

\shorttitle{Twisting, reconnecting magnetospheres}
\shortauthors{K. Parfrey, A. M. Beloborodov, and L. Hui}

\begin{document}

\title{Twisting, reconnecting magnetospheres and magnetar spindown}
\author{Kyle Parfrey\altaffilmark{1,2}, Andrei M. Beloborodov\altaffilmark{3}, and Lam Hui\altaffilmark{2,3}}
\affil{
$^{1}$Astronomy Department, Columbia University, 550 West 120th Street, New York, NY 10027, USA; kyle@astro.columbia.edu\\
$^{2}$Institute for Strings, Cosmology, and Astroparticle Physics (ISCAP), Columbia University, New York, NY 10027, USA\\
$^{3}$Physics Department and Columbia Astrophysics Laboratory, Columbia University, 538 West 120th Street, New York, NY 10027, USA
}

\label{firstpage}
\begin{abstract}
We present the first simulations of evolving, strongly twisted magnetar magnetospheres. Slow shearing of the magnetar crust is seen to lead to a series of magnetospheric expansion and reconnection events, corresponding to X-ray flares and bursts. The axisymmetric simulations include rotation of the neutron star and the magnetic wind through the light cylinder. We study how the increasing twist affects the spindown rate of the star, finding that a dramatic increase in spindown occurs. Particularly spectacular are explosive events caused by the sudden opening of large amounts of overtwisted magnetic flux, which may be associated with the observed giant flares. These events are accompanied by a short period of ultra-strong spindown, resulting in an abrupt increase in spin period, such as was observed in the giant flare of SGR~1900+14.
\end{abstract}


\section{Introduction}

Magnetars are neutron stars with luminosity powered by the dissipation of magnetic energy \citep{1992ApJ...392L...9D}. Their magnetospheres evolve on timescales much shorter than the age of the star. The magnetosphere is attached to the stellar crust and is thought to be deformed by crustal motions (either gradual or catastrophic). 

The crust is practically incompressible, and its motion must be pure shear. Shearing of the magnetospheric footpoints generates external twists ($\nabla\times\bB\neq 0$) and hence excites external electric currents. Ohmic dissipation of these currents generates non-thermal activity \citep{2000ApJ...543..340T, 2002ApJ...574..332T, 2012arXiv1201.0664B}. A quasi-steady twist implanted in the magnetosphere is erased on the Ohmic timescale $\sim 1$~yr, as it gradually dissipates via a continual electric discharge \citep{2007ApJ...657..967B, 2009ApJ...703.1044B}. Thus, the crustal shearing creating the twist must be sufficiently fast, with vorticity $\omega_c\simgt 1$~rad~yr$^{-1}$.
    
Observed magnetars rotate with periods $P=2-12$~s and are 
gradually spun down, on a timescale comparable to the age of the star,
$t\sim 10^4$~yr \citep[e.g.][]{2008A&ARv..15..225M}.
The spindown torque acting on the star is controlled by the open magnetic 
flux passing through the light cylinder of radius $\RLC=cP/2\pi$.
As the magnetosphere is twisted, its energy density grows and it tends to 
inflate \citep[e.g.][]{1995ApJ...443..810W}; this opens more magnetic flux, increasing
the spindown torque. Thus, the 
magnetic twists, besides generating non-thermal emission, can also account for
the observed temporal variations in spindown rate $\dot{P}$.

For example, an outburst from XTE~J1810-197 was followed by a change in $\dot{P}$
of more than a factor of three over nine months \citep{2007ApJ...663..497C}. The spindown rates of SGRs 1900+14 and 1806-20 have been observed to vary by a factor of four over timescales of months \citep{2002ApJ...576..381W}. The 27 August 1998 giant flare from SGR 1900+14 was coincident with a fractional increase in the spin period of $\Delta P/P = 10^{-4}$ \citep{1999ApJ...524L..55W}. Because of an 80 day gap in observations before the flare, the behavior of $\dot{P}$ is unknown. It is unclear if there was a gradual change in spindown over this period, if $\Delta P$ resulted from a brief and dramatic increase in magnetic torque, or if it was a result of a sudden change inside the neutron star, a sort of ``anti-glitch'' \citep{2000ApJ...543..340T}.

In a simple axisymmetric model, the twist amplitude $\psi$ is measured by the
azimuthal angle between the magnetospheric footpoints.
Twists of small amplitude $\psi\ll 1$ change $\dot{P}$
by a small fraction $\sim\psi^2$ \citep{2009ApJ...703.1044B}. 
Significant changes in $\dot{P}$ are associated with large $\psi\sim 1$.
On the other hand, when $\psi$ exceeds a  critical value $\sim 1$, the
twisted magnetic equilibrium loses stability \citep[e.g.][]{2002ApJ...574.1011U}.
The instability is similar to coronal mass ejections from the solar corona---part of the 
magnetic energy is ejected, and the twist amplitude is reduced.

The purpose of this paper is to study the response of 
the magnetosphere of a rotating star to strong twisting. 
We present the first numerical simulations of this problem, in its simplest, 
axially symmetric version. We consider the aligned rotator 
(a dipole magnetosphere whose axis is parallel to the spin axis of the star);
the twisting motion of the stellar crust is modeled as a slow rotation
of either one polar cap, or a band of latitudes, with respect to the rest of the star.

The magnetospheric field lines are conveniently labeled with the poloidal 
flux function $f$; for the twisted dipole configuration $f=2\pi\mu \sin^2\theta/\rstar$, 
       where $\mu$ is the magnetic
       dipole moment of the star, $\rstar$ is the stellar radius, and $\theta$ 
       is the colatitude of a field line's northern footpoint.
The gradual pumping of the twist is described by \citep[see][]{2011heep.conf..299B}
\begin{equation}
\label{eq:pump}
   \dot{\psi}(f)=\omega_c(f)+2\pi\,c\, \frac{\partial \Phi}{\partial f}\equiv\omega(f).
\end{equation} 
Here $\omega_c=\dot{\phi}_n-\dot{\phi}_s$ is the differential angular velocity 
of the moving crust (``n'' and ``s'' refer to the northern and 
southern footpoints of the field line), 
and $\Phi$ is the voltage between the footpoints.
The resistive term $2\pi c\, \partial \Phi/\partial f$ is negligible if $\omega_c \gg 1\, {\rm rad\; yr^{-1}}$. When this term is important, it can temporarily amplify the twist near the axis while reducing it everywhere else \citep{2009ApJ...703.1044B}.
The approximate effect of the resistive term can be incorporated into an ideal calculation by attributing the entire effective $\omega$ to footpoint shearing. This is the approach we take in our simulations, where the magnetosphere is described as ideal and force-free (except in current sheets).

Our simulations are novel in at least two respects. Firstly, previous time-dependent numerical investigations of sheared magnetospheres have used non-relativistic MHD \citep{1994ApJ...430..898M, 1999ApJ...510..485A}; we solve the equations of force-free electrodynamics, the appropriate limit for plasmas in ultra-strong magnetic fields. In this regime, the plasma can be approximated as massless. Secondly, our simulations include stellar rotation, with the light cylinder radius within the computational domain.

\section{Numerical simulations}

Our simulations are performed with \textsc{PHAEDRA}, a parallel pseudospectral code for force-free electrodynamics \citep{2012MNRAS.tmp.2963P}. The equations solved are Maxwell's equations with a non-linear current density $\J$, which incorporates the plasma response:
\begin{equation}
\dd_t \B = - \curlE , \qquad   \dd_t \E = \curlB - 4\pi\J,
\label{eq:maxwell}
\end{equation}
\beq
\J = \frac{\vec{B} \cdot \curlB - \vec{E} \cdot \curlE }{4\pi B^2} \B + \divE \, \frac{\vec{E}\times\vec{B}}{4\pi B^2}.
\label{eq:current}
\eeq
These equations ensure that the Lorentz force vanishes, $(\divE)\E/4\pi + \J\times\B=0$, and that $\E$ is perpendicular to both $\B$ and $\J$, implying zero Ohmic dissipation.
Motion of the stellar surface is effected by imposing an electric field at the inner boundary of the domain, which is induced as the 
highly-conducting crust drags the frozen-in magnetic field, 
$\E = - \ls \lp\vec{\Omega}+\vec{\omega}\rp\times\vec{r} \rs \times \B$,
where $\vec{\Omega}=2\pi\,\vec{\rm e_z}/P$ represents rotation and $\vec{\omega}$ represents shearing of the star.
The outer boundary condition (at $r\approx 3\RLC$) allows free escape
of Poynting flux.
The code has very low unphysical numerical diffusion or dissipation on resolved scales, while a high-order spectral filtering procedure applies some controlled dissipation near the grid scale, which becomes the resistive length scale at which reconnection can occur. This is desirable, because the magnetosphere should be very close to ideal, except in current sheets where reconnection is expected. For our purposes it is important only that resistivity be confined to current sheets, and that these can become as thin as possible---it is irrelevant that it is introduced with a filter rather than an explicit term in the equations of motion.

In our axisymmetric simulations the magnetic axis is parallel to both the rotation and twist angular velocity vectors, $\vec{\Omega}$ and $\vec{\omega}$. Each simulation begins with the stationary solution for a rotating star with no twisting, $\vec{\omega}=0$ (Fig.~1a). It is found by setting a dipole magnetic field into rotation and allowing it to relax to a steady state, as described in \citet{2012MNRAS.tmp.2963P}. 
Then $\vec{\omega}$ is superimposed, so that either a northern polar cap, or a ring of northern latitudes, rotates slightly faster than the rest of the star. The polar cap, extending down to the colatitude $\thpc$, is modeled by the twisting profile
\beq
\omega_{\rm cap}(\theta) =  \frac{\omega_{0}}{1 + \exp \ls \kappa \lp \theta - \thpc\rp \rs},
\eeq
and the ring profile is
\beq
\omega_{\rm ring}(\theta) =  \frac{\omega_{0}}{1 + \exp \lb \kappa \ls  \left | \theta - \theta_{\rm ctr} \right | - \Delta \rs \rb},
\eeq
where $\theta_{\rm ctr}$ is the colatitude of the center of the ring and $\Delta$ is its angular half-width. We set $\kappa = 50$ and $\omega_0 = \Omega/200$; the twisting rate is approximately constant within the twisted region, and drops quickly to zero elsewhere. 

The twisted polar cap is bigger than the footprint of the open flux 
bundle of the (untwisted) rotating dipole, 
$\thpc>\throt$, where $\throt\approx(\rstar/\RLC)^{1/2}$. 
The ratio of magnetic fluxes emerging through the caps $\theta<\thpc$
and $\theta<\throt$ defines a characteristic dimensionless parameter, 
\beq
a = \frac{\ftw}{\frot}=\frac{\sin^2\thpc}{\sin^2\throt} > 1.
\label{eq:a}
\eeq
A similar dimensionless parameter can be used to label the flux surfaces that bound the ring twisting region. A ring defined by the flux surfaces $a_1$ and $a_2$ extends from $\theta_1 = \theta_{\rm ctr} - \Delta$ to $\theta_2 = \theta_{\rm ctr} + \Delta$, where $\sin^2 \theta_1 = a_1 \sin^2\throt$ and $\sin^2 \theta_2 = a_2 \sin^2\throt$; the magnetic flux through the twisted ring is $(a_2-a_1) \frot$.

\begin{deluxetable}{c c c c c}
\tablewidth{80mm}
\tablecaption{Simulation parameters}
\tablehead{
\colhead{Run} & \colhead{$R_{\rm LC}/\rstar$} & \colhead{Twist Profile} & \colhead{$a$} &
  \colhead{$N_r \times N_\theta $}
}
\startdata
A & 20 & Polar Cap & 6 &  $384 \times 255$ \\ 
B & 40 & Polar Cap & 12 & $768 \times 507$ \\ 
C & 40 & Ring & 9--12 &  $768 \times 507$
\enddata
\end{deluxetable}

\begin{figure*}
\centering
\includegraphics[width=172mm]{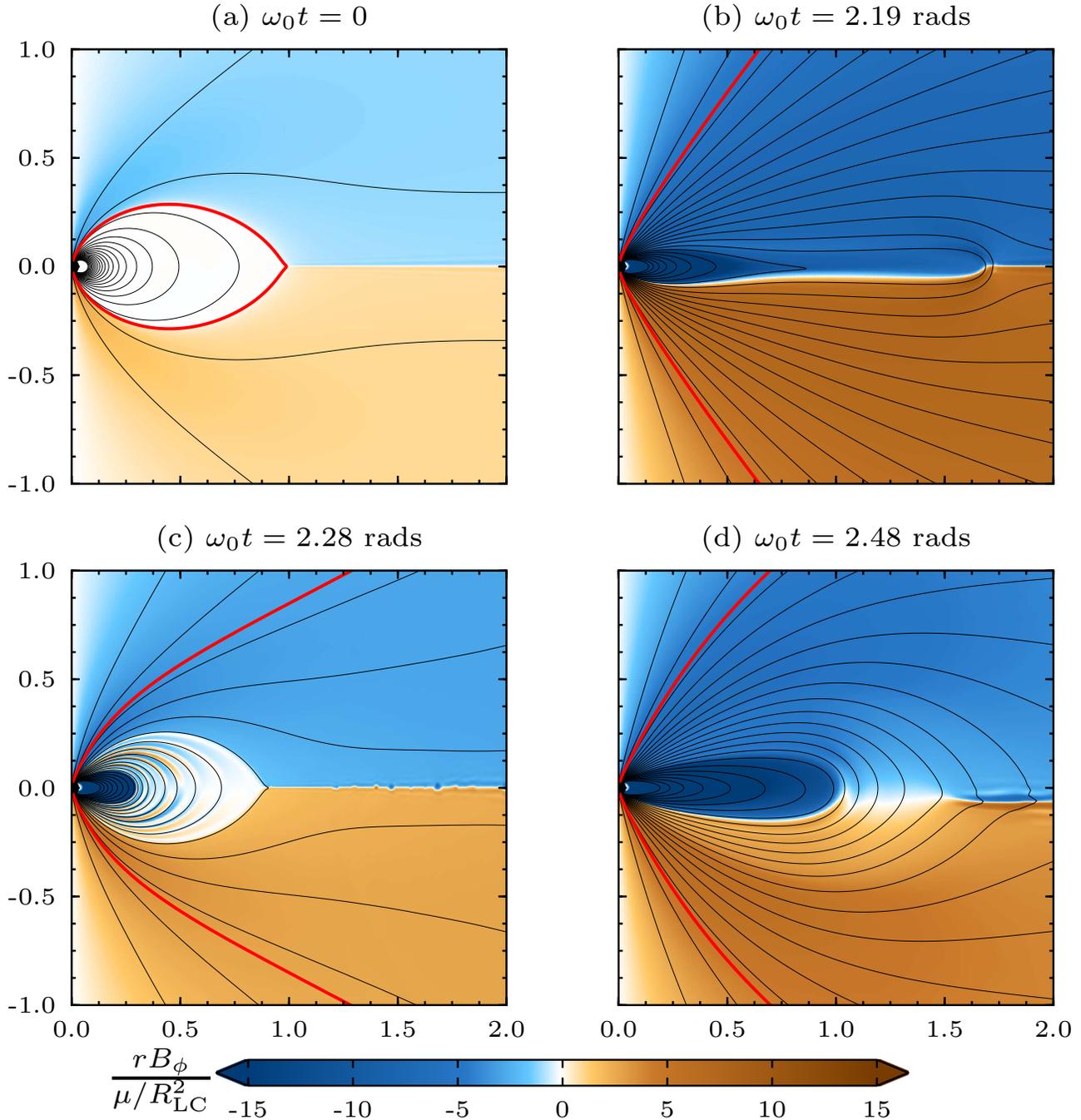}
\caption{\label{fig1} The first two expansion and reconnection events from run B. 
Color shows toroidal magnetic field times the radial coordinate, in units where this quantity is initially $\sim 1$ at the light cylinder. 
Black curves show twenty-two poloidal field lines, equally spaced in poloidal flux function between $f = 0.01 f_{\rm max}$ and $(1/3) f_{\rm max}$, where $f_{\rm max} = 2\pi\mu/\rstar$;
one additional field line, that initially closes at the light cylinder, is shown in red.  Axes are labeled in units of $\rlc = 40\,\rstar$.
(a) The initial state (untwisted rotating star); 
(b) the last outward ``breathing'' motion before the first reconnection; (c) after the first reconnection, the inner dark blue region is the twisted reservoir, the almost white area is the cavity; (d) detonation---the reservoir expands explosively.}
\end{figure*}

Three simulations A, B, and C are detailed in Table~1, where $N_r$ and $N_\theta$ are the numbers of grid points in the radial and meridional directions. In all simulations, the magnetosphere is twisted up to a maximum angle of $\omega_0\tmax = 8$ rads over 263 rotation periods. Fig.~1 illustrates the behavior of run~B.

For runs A and B, the evolution proceeds as follows. As the twisting begins, toroidal magnetic field builds up on the closed field lines connected to the sheared cap (on open field lines the twist is emitted to infinity rather than accumulated). As the twist pumping $\dot{\psi}$ is slow compared with the Alfv\'en crossing time, the magnetosphere first evolves through a sequence of quasi-equilibrium states.
The increased magnetic pressure in the closed zone causes these field lines to expand outwards, pushing more flux through the light cylinder and hence increasing the spindown torque on the star. 
The newly-opened field lines lose their twist---it is emitted to infinity---and establish the usual spindown-wind configuration where the field changes sign across the equatorial plane. The magnetic field discontinuity is supported by the equatorial current sheet, terminating at a Y-point which separates closed and open flux. The magnetic field remains almost reflection-symmetric about the equator (modulo differences in sign between oppositely directed flux), because the twisting timescale is much longer than the wave-crossing timescale on the closed field lines, allowing waves to distribute the twist nearly symmetrically. 

The expansion rate increases while the twist $\psi$ grows with constant rate $\omega$, and eventually the field lines expand through the light cylinder faster than they adjust to a new spindown steady state; this sets up a pattern of magnetospheric ``breathing,'' where the amount of open flux oscillates, via reconnection in the current sheet, with increasing amplitude and a period of $\sim 7\; R_{\rm LC}/c$ (Fig.~1b).

After a significant fraction of the polar cap flux has been opened, at  $\psi \sim 2$, the  recently opened field lines become unstable to catastrophic reconnection and a large fraction of the open flux reconnects, bringing the Y-point within the light cylinder. The released magnetic energy is expelled in a plasmoid-fragmented outflow in the equatorial plane. The Y-point quickly moves back to the light cylinder, opening up a cavity of zero toroidal field between the low-lying strongly twisted region and the open flux bundle (Fig.~1c). We will term the evolution from Fig.~1a to Fig.~1c a ``gradual'' expansion and reconnection event.

The accumulated twist $\psi=\omega_0 t$ remains intact on those lines, lying closer to the star, which never opened; let us call these field lines the ``twisted reservoir.'' 
As the shear motion of the polar cap continues, 
toroidal field is added to 
both the twisted reservoir and the cavity. Before a new
significant $\psi$ builds up in the cavity, the over-twisted reservoir
becomes unstable and expands outward explosively, and all the overlying field lines are briefly pushed through the light cylinder (Fig.~1d). This process is reminiscent of magnetic detonation \citep{1997PhR...283..185C}. There is a narrow spike in the spindown torque, larger than in the previous event because more flux is opened. The enhanced open flux immediately reconnects and closes, leaving a reservoir of lesser volume with $\psi\approx \omega_0 t$ and a bigger cavity with $\psi\approx 0$. This is an ``explosive'' expansion and reconnection event.

\begin{figure}
\begin{center}
\includegraphics[width=85mm]{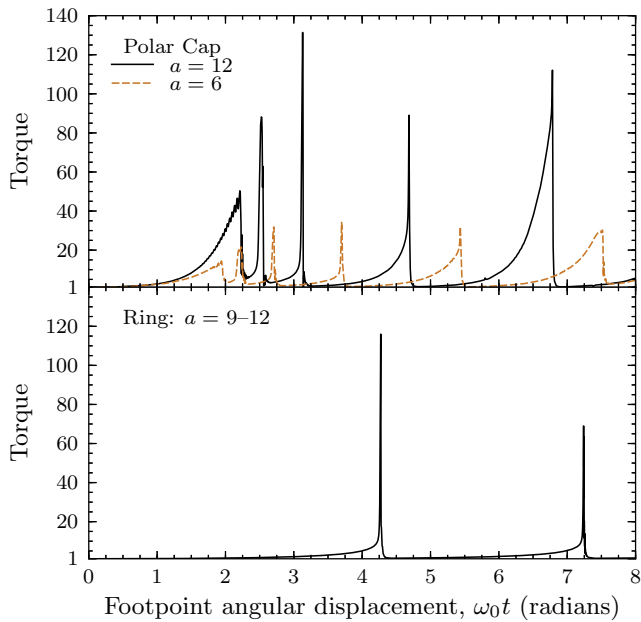}
\caption{\label{fig2} Spindown torque on the star (integrated angular momentum flux through the stellar surface), in units of the torque on the untwisted rotating star, versus the relative angular displacement of the twisted region, $\omega_0 t$. Runs A and B (polar caps) are in the upper panel, run C (ring profile) is in the lower panel.}
\end{center}
\end{figure}

This cycle of a twisted reservoir exploding and reconnecting is repeated several times as the shearing continues.
In the first few events successively more flux is opened, leading to higher spikes in spindown rate, and a deeper trough in torque immediately following reconnection, since the twisted reservoir comprises ever fewer field lines. The spikes in spindown power become increasingly narrow as they become more energetic, with the most dramatic events having a duration of only a few light crossing times of the light cylinder (Fig.~\ref{fig2}).

The time between events grows, because the twisted reservoir is shrinking. Eventually, the additional twist needed to detonate the reservoir
becomes greater than $\sim 2$ rads, and the cavity fills with enough toroidal field to initiate a gradual reconnection event before the reservoir explodes. After this point the picture becomes more complicated, as there can be a twist-free cavity on top of a twisted reservoir comprising both flux that did not open in the last event, and flux that has never opened. Different amounts of flux will then expand and open depending on the history of the system.

The heights of the torque spikes scale approximately as $a^2$, 
as can be seen in Fig.~\ref{fig2}. This scaling, which is confirmed by further simulations
with $a$ from 2 to 12, is expected. The spindown luminosity $L$ is related to the magnetic flux through the light cylinder $f_{\rm LC}$ by
\beq
L \approx \frac{1}{c} \lp\frac{f_{\rm LC}}{P}\rp^2,
\label{eq:Lfromf}
\eeq
where $P$ is the rotation period \citep{Goldreich:1969p4440}. The opening of the twisted, inflated flux bundle increases $f_{\rm LC}$ by nearly a factor of $a$, $f_{\rm LC} \approx a f_{\rm rot}$, which boosts $L$ by a factor of $a^2$. When averaged over the entire run time of the simulations, the spindown power in runs A and B is enhanced by factors of 5.45 and 9.98 respectively; thus long-term average spindown is linear in $a$.

Run C evolves slightly differently, because only deep closed flux is twisted. There is no gradual opening---all events are explosive, and require a larger twist angle than in the polar cap simulations before the accumulated magnetic pressure is great enough to overcome the restraining tension of the overlying untwisted field lines. The torque peaks scale as the square of the total opened flux (equation~\ref{eq:Lfromf}), which includes both the twisted ring and the overlying magnetosphere, because during an event the expanding twisted field lines push the overlying untwisted flux through the light cylinder. The average spindown power is enhanced by a factor of 2.54.

\section{Discussion}

We find that the twisted magnetosphere undergoes a succession of opening and reconnection events, accompanied by large increases in the spindown power (Fig.~\ref{fig2}). There are two main classes of field expansion and reconnection events. 
(1) The gradual kind involves smooth expansion of the magnetosphere through the light cylinder; it causes spindown to increase on the twisting timescale $\omega^{-1}$ until sudden reconnection restores the normal regime.
(2) Explosive events are caused by the sudden breakout of flux under high magnetic pressure, whose expansion had been slowed by the overlying  magnetosphere; these may be as brief as a few light crossing times of the light cylinder, which is comparable to the spin period of the star.
In both cases the reconnection, and hence energy dissipation, phase 
is very short (as can be seen in the near-vertical downstrokes in Fig.~\ref{fig2}). 
 
The energy stored by twisting a polar cap $\thpc$ of a star with 
magnetic moment $\mu$ and radius $\rstar$ is given by \citep{2009ApJ...703.1044B},
\beq
\label{eq:Etw}
    \Etw\approx \frac{\mu^2 \psi^2 \sin^6\!\thpc}{24 \rstar^3}
        \sim 4\times10^{46} \mu_{33}^2 \psi^2 \sin^6\!\thpc {\rm ~erg},
\eeq
where $\mu_{33} = \mu/10^{33}$ G cm$^3$ and $\rstar \approx 10^6$ cm.
During an explosive event, this energy is partially released in a magnetic outflow and partially radiated
in a powerful flare. Our simulations suggest that such events are accompanied
by a brief increase in spindown torque by a factor as large as $a^2$ (see equation~\ref{eq:Lfromf}), which 
leads to an abrupt increase in spin period. 
Then the fractional change in period is approximately given by 
\beq
   \frac{\Delta P}{P}\sim a^2 \frac{\Delta t}{t_0}\sim 
   a^2 \left(\frac{c\, \Delta t}{2\pi R_{\rm LC}}\right) \dot{P}_0,
\eeq
where $\dot{P}_0=P/t_0$ is the spindown rate for the untwisted magnetosphere
and $\Delta t$ is the duration of the torque increase by $a^2$.
A typical $\thpc\sim$ 0.3--0.6, which is sufficient for observed giant flares
(equation~\ref{eq:Etw}), corresponds to $a\sim 3\times 10^3$ (equation~\ref{eq:a}). 
SGR~1900+14 has $\dot{P}_0 \approx 10^{-10}\, \rm{s\,\, s}^{-1}$, so a short-duration 
spike in spindown, with $\Delta t $ comparable to $R_{\rm LC}/c$, 
can give $\Delta P/P\sim 10^{-4}$ as observed for the August 1998 flare.
This suggests that huge anti-glitches may be explained without 
recourse to sudden changes in the stellar interior.
For the December 2004 flare in SGR~1806-20, which was 
two orders of magnitude more energetic, one could expect even larger 
$\Delta P$, which is not observed---the measured upper limit for $\Delta P/P$ is 
$5\times 10^{-6}$ \citep{2007ApJ...654..470W}. 
Note that $\Delta P\propto a^2$ while $\Etw \propto a^3$. Variations in $a$, $\mu$, and twist geometry may lead to significant variations in $\Delta P$. 

The average $\dot{P}$ in the twisted-cap model is increased by a factor of $a$ (\S~2), 
suggesting that the use of the standard dipole estimate for $\dot{P}$ 
may over-estimate the magnetic field.
This effect is weaker if a ring is twisted instead 
of a polar cap---then the explosive mode of field opening is dominant and
the average $\dot{P}$ is not increased as much.

We stress that explosive opening of the magnetosphere is not limited to certain twist profiles. Even in the case of a polar cap twisted all the way to the magnetic axis, when gradual collective opening would be expected, explosive events occur after the system has reconnected at least once. This is because those field lines which did not reconnect behave like a ring of twisted flux, and this twisted reservoir can become unstable before appreciable twist accumulates on the overlying field lines. 

The effect of relaxing axial symmetry remains to be seen. New global instabilities would become possible, in particular kink instability of the inflated magnetosphere. We plan to investigate this possibility with future simulations.

This work was supported in part by NASA (NNX-10-AI72G and NNX-10-AN14G), and the DOE (DE-FG02-92-ER40699).


\begin{thebibliography}{21}

\bibitem[{{Antiochos} {et~al.}(1999){Antiochos}, {DeVore}, \&
  {Klimchuk}}]{1999ApJ...510..485A}
{Antiochos}, S.~K., {DeVore}, C.~R., \& {Klimchuk}, J.~A. 1999, \apj, 510, 485

\bibitem[{{Beloborodov}(2009)}]{2009ApJ...703.1044B}
{Beloborodov}, A.~M. 2009, \apj, 703, 1044

\bibitem[{{Beloborodov}(2011)}]{2011heep.conf..299B}
{Beloborodov}, A.~M. 2011, in High-Energy Emission from Pulsars and their
  Systems, ed. {D.~F.~Torres \& N.~Rea}, 299

\bibitem[{{Beloborodov}(2012)}]{2012arXiv1201.0664B}
{Beloborodov}, A.~M. 2012, ArXiv: 1201.0664

\bibitem[{{Beloborodov} \& {Thompson}(2007)}]{2007ApJ...657..967B}
{Beloborodov}, A.~M., \& {Thompson}, C. 2007, \apj, 657, 967

\bibitem[{{Camilo} {et~al.}(2007)}]{2007ApJ...663..497C}
{Camilo}, F., {et~al.} 2007, \apj, 663, 497

\bibitem[{{Cowley} \& {Artun}(1997)}]{1997PhR...283..185C}
{Cowley}, S.~C., \& {Artun}, M. 1997, Phys. Rep., 283, 185

\bibitem[{{Duncan} \& {Thompson}(1992)}]{1992ApJ...392L...9D}
{Duncan}, R.~C., \& {Thompson}, C. 1992, \apjl, 392, L9

\bibitem[{Goldreich \& Julian(1969)}]{Goldreich:1969p4440}
Goldreich P., Julian W.~H., 1969, \apj, 157, 869

\bibitem[{{Mereghetti}(2008)}]{2008A&ARv..15..225M}
{Mereghetti}, S. 2008, \aapr, 15, 225

\bibitem[{{Mikic} \& {Linker}(1994)}]{1994ApJ...430..898M}
{Mikic}, Z., \& {Linker}, J.~A. 1994, \apj, 430, 898

\bibitem[{{Palmer} {et~al.}(2005)}]{2005Natur.434.1107P}
{Palmer}, D.~M., {et~al.} 2005, Nature, 434, 1107


\bibitem[{{Parfrey} {et~al.}(2012){Parfrey}, {Beloborodov}, \&
  {Hui}}]{2012MNRAS.tmp.2963P}
{Parfrey}, K., {Beloborodov}, A.~M., \& {Hui}, L. 2012, \mnras, 2963

\bibitem[{{Thompson} {et~al.}(2000){Thompson}, {Duncan}, {Woods},
  {Kouveliotou}, {Finger}, \& {van Paradijs}}]{2000ApJ...543..340T}
{Thompson}, C., {Duncan}, R.~C., {Woods}, P.~M., {et~al.} 2000, \apj, 543, 340

\bibitem[{{Thompson} {et~al.}(2002){Thompson}, {Lyutikov}, \&
  {Kulkarni}}]{2002ApJ...574..332T}
{Thompson}, C., {Lyutikov}, M., \& {Kulkarni}, S.~R. 2002, \apj, 574, 332

\bibitem[{{Uzdensky}(2002)}]{2002ApJ...574.1011U}
{Uzdensky}, D.~A. 2002, \apj, 574, 1011

\bibitem[{{Wolfson}(1995)}]{1995ApJ...443..810W}
{Wolfson}, R. 1995, \apj, 443, 810

\bibitem[{{Woods} {et~al.}(1999)}]{1999ApJ...524L..55W}
{Woods}, P.~M., {et~al.} 1999, \apjl, 524, L55

\bibitem[{{Woods} {et~al.}(2002)}]{2002ApJ...576..381W}
{Woods}, P.~M., {et~al.} 2002, \apj, 576, 381

\bibitem[{{Woods} {et~al.}(2007){Woods}, {Kouveliotou}, {Finger}, {G{\"o}{\u
  g}{\"u}{\c s}}, {Wilson}, {Patel}, {Hurley}, \&
  {Swank}}]{2007ApJ...654..470W}
{Woods}, P.~M., {et~al.} 2007,
  \apj, 654, 470
  
\end{thebibliography}
\end{document}